\documentclass[amsmath,amssymb,aps,twocolumn,jcp,floatfix]{revtex4}

\usepackage{mathrsfs}
\usepackage{amsmath}
\usepackage{amssymb}
\usepackage{color}
\usepackage{graphicx}	
\usepackage{bm}
\usepackage{ulem} 
\usepackage{titlesec}
\titleformat{\paragraph}[runin]
{\bfseries\scshape}{\theparagraph}{1em}{}
\usepackage{braket}
\usepackage{multirow}

\usepackage[printwatermark]{xwatermark}
\usepackage{xcolor}

\newcommand{\be}{\begin{equation}}
\newcommand{\ee}{\end{equation}}
\newcommand{\bef}{\begin{figure}}
\newcommand{\eef}{\end{figure}}
\newcommand{\bea}{\begin{eqnarray}}
\newcommand{\eea}{\end{eqnarray}}

\begin{document}
\title{Depletion induced demixing and crystallization in binary colloids subjected to an external potential barrier}
\author{Mahammad Mustakim and A. V. Anil Kumar}
\thanks{Corresponding author: \texttt{anil@niser.ac.in}.}
\affiliation{School of Physical Sciences, National Institute of Science Education and Research, HBNI, Bhubaneswar-752050, India}

\date{\today}
\begin{abstract}
A binary colloidal mixture of unequal sizes, subjected to an external potential barrier, has been investigated using canonical ensemble molecular dynamics simulations. 
The attractive depletion interaction between the external barrier and larger species in the binary mixture causes the mixture to phase separate. At higher volume fractions,
a pure phase of larger particles forms near the potential barrier, and the local density of this pure phase is high enough that a face centered cubic crystalline domain is formed at this region. This crystalline phase diffuses perpendicular to the external potential barrier. The temperature dependence of diffusivity of larger particles is
non-Arrhenius and changes from sub-Arrhenius to super-Arrhenius as the volume fraction increases. This crossover from sub-Arrhenius to super-Arrhenius diffusion coincides with the crystalline formation near the potential barrier.

\end{abstract}
\maketitle

\section{Introduction}

Suspensions of colloidal particles exhibit similar phase behaviors as atomic systems\cite{pusey1,asherie}. This atom-colloid analogy makes the colloids an important model system to study the phase transitions. The solvent mediated effective interactions between the colloidal particles determine the physical properties
of a colloidal dispersion and these effective interactions can be tuned to have different ranges as well as strengths\cite{yethiraj}. For example, the interaction can be long-ranged repulsive on one end, while it can be tuned to nearly hard-spheres at the other end. This tunability in interactions enables us to use colloids to study a wide range of phase behaviors and transitions. Crystallization and melting are among the most studied phenomena both from fundamental 
as well as technological points of view. In particular, colloidal crystals are of great interest for developing new novel materials such as optical fibers\cite{park,luis}, 
photonic bandgap materials\cite{hynninen,wan,harini} etc. The volume fraction is one of the most important factors in determining the phase behavior of all the colloidal suspensions. 
For hard-sphere colloids, the volume fraction is the only controlling parameter that determines whether the suspension is in a fluid state or in a solid state. 
The transition from a fluid state to a crystalline state occurs at higher volume fractions($\phi$ = 0.49 for hard sphere colloids). 
In colloidal systems with long-ranged 
repulsive interactions, crystal phases can be formed at lower volume fractions. These crystals are body centered cubic crystals, named 'Wigner crystals'. However, when the interactions are short ranged, the colloidal suspensions tend to be in a liquid state at lower volume fractions. Crystallization at low volume fractions
can be achieved by manipulating the local concentration of colloids by dielectrophoresis\cite{sullivan1,sullivan2} and diffusiophoresis\cite{rainmuller}.
It will be interesting to investigate how the local concentration can be modified making use of the entropic effects such that the colloidal systems with short ranged 
interactions form crystals at lower volume fractions.

The binary mixture of colloids, where the components differ in their sizes, shows rich phase behavior as well as dynamical properties due to the excluded volume effects. 
The disparity in the sizes of components entropically favors an effective attraction between the larger particles. This entropically driven effective attraction 
between the larger($l$) components in the binary mixture is called depletion interactions\cite{asakura,vrij} and leads to interesting structural and dynamical 
properties, especially crystallization, gelation, and self-assembly of colloidal particles\cite{kim, pham, walz}. Recently it has been shown that when a binary colloidal mixture is exposed to an external repulsive potential, there exists
an attractive depletion interaction between the potential barrier and the larger components in the mixture.
This depletion interaction significantly alters the structural and dynamical properties of the binary mixture
and exhibits very interesting behaviors\cite{anil,anil1,mustakim,jalim}. For example, it favors the demixing of the binary mixture and an $l$-rich phase is formed near the external potential barrier\cite{anil}. Also, there have been interesting observations about the dynamics of both components in the mixture. 
The smaller($s$) particles exhibit a slowing down of dynamics at lower temperatures like in the case of supercooled liquids, even though the volume fraction is quite low\cite{anil1}. Meanwhile, the larger particles continue to undergo normal diffusion even at low temperatures. The temperature dependence of diffusion is also interesting. The smaller particles follow an Arrhenius behavior even though their dynamics has slowed down. However, the diffusion of larger particles deviates from Arrhenius behavior and shows a sub-Arrhenius temperature dependence\cite{mustakim}. All these results are obtained at a total volume fraction of $\phi$ = 0.2. 

In this article, we extend these investigations to higher volume fractions using molecular dynamics simulations. The effect of volume fraction on the structural 
and dynamical properties have been studied. 
The demixing becomes stronger as volume fraction increases and the $l$-rich phase forms crystals of larger particles about $\phi$=0.40. This crystalline phase, 
which is formed at the region of the external potential barrier moves perpendicular to the barrier with a non-zero diffusion coefficient. This seems to be surprising since the local volume fraction is large enough so that the particle motion is very much hindered. Such moving crystals have been observed earlier in the nonequilibrium systems, where the crystals are subjected to heating or cooling or exposed to light\cite{panda1,panda2,shima,commins}. However, in our model system, there are no driving forces present and the whole system is in equilibrium. It has been also found that the temperature dependence of the diffusion of larger particles along the direction of external potential changes from sub-Arrhenius to super-Arrhenius when the crystal formation occurs in the system. 

The paper is organized as follows. In section II, we describe our model system and the simulation techniques and outline the parameters used in the simulations. 
In section III, we present the structural and dynamical properties of the system. Here we discuss the important observations of crystalline phase formation as well as the diffusion of this crystalline phase. We also discuss the variation from Arrhenius behavior in the diffusion of larger particles and the crossover from sub-Arrhenius to super-Arrhenius diffusion. Finally, the work is summarized in section IV.

\begin{figure}
\centering
\includegraphics[width=6cm, angle=-0]{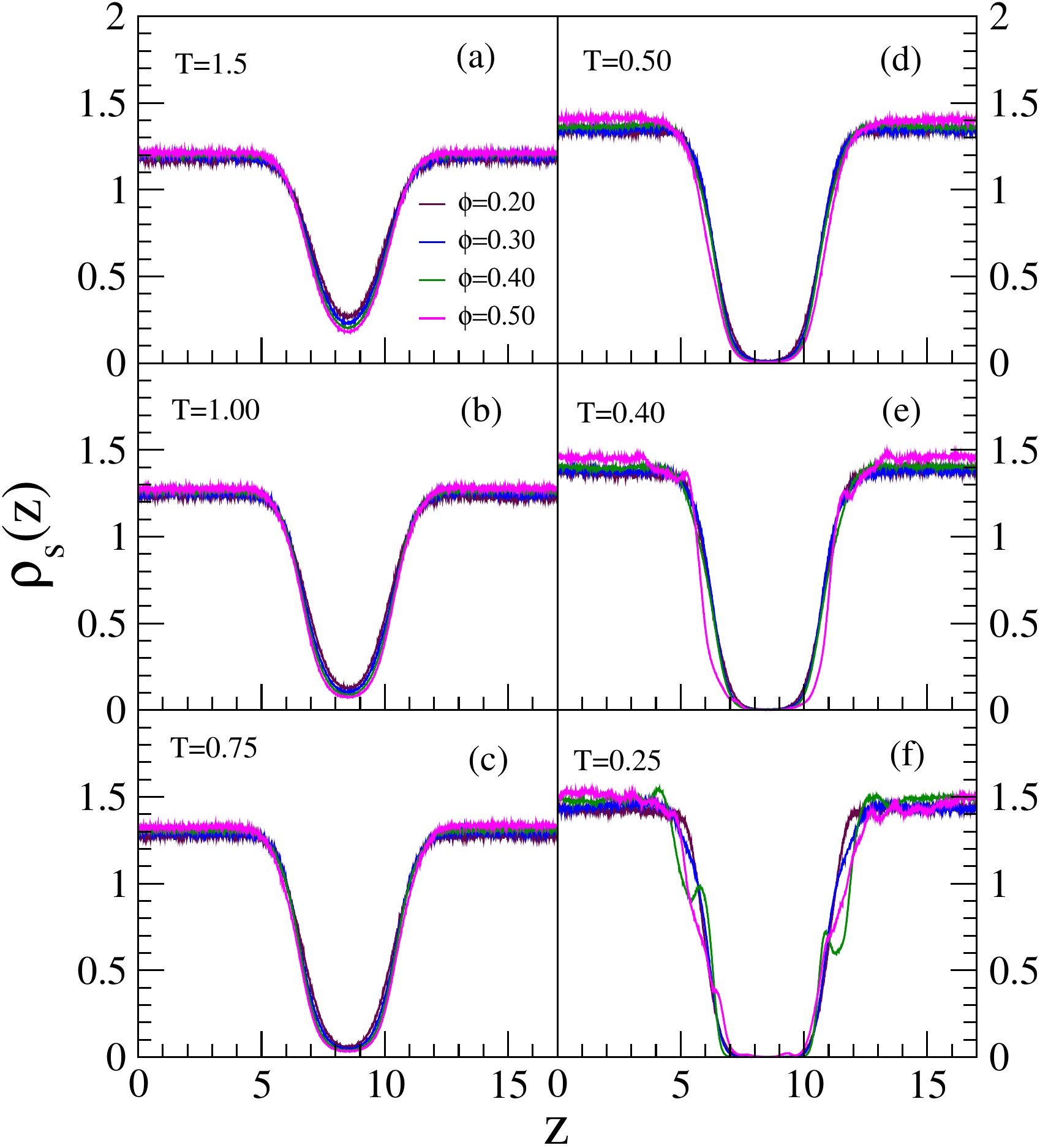}
\caption{Density profile of smaller particles at four different volume fractions $\phi$ = 0.20, 0.30, 0.40 and 0.50 and at different temperatures (a) $T$=1.5, (b) $T$ = 1.0, (c) $T$ = 0.75, (d) $T$ = 0.50, (e) $T$ = 0.40 and (e) $T$ = 0.25.}
\label{densmall}
\end{figure}

\section{Model and simulation details}
We have carried out the canonical molecular dynamics simulations on a system of binary mixture of colloidal particles subjected to an external repulsive potential. 
The system contains two different sized particles, interacting via a soft sphere repulsion which is given by
\begin{equation}
 V_{ab}(r_{ij}) = \epsilon_{ab} \Big(\frac{\sigma_{ab}}{r_{ij}}\Big)^{12}
\end{equation}

\noindent where $r_{ij}$ is the inter-particle distance and ($a, b) \in (l(large), s(small)$. The masses of two species are kept same since we do not 
want the effect of mass difference to be coupled with the effect of depletion interaction in the dynamics of the binary mixture.
We take the simulation box of length, $L=17$, and periodic boundary 
conditions are applied in all the three spatial directions. The system is subjected to an external repulsive barrier of Gaussian form at the center 
of the simulation box along $z-$ direction only\cite{anil},
\begin{equation}
V_{ext}(z) = \epsilon_{ext} \,\, e^{-\Big(\frac{z-z_0}{w}\Big)^2}
\end{equation}

\noindent The width and height of the barrier are $w$ = 3.0 and $\epsilon_{ext}$ = 3.0 respectively. The details of the model parameters and reduced units are 
same as in Ref. \cite{anil}. We take an equivolume mixture of large and small colloidal particles and carried out simulations for different total volume 
fractions of $\phi=0.20$, $\phi=0.25$, $\phi=0.30$ $\phi=0.35$, $\phi=0.40$, $\phi=0.425$, $\phi=0.45$, and $\phi=0.50$ to observe the changes in the structure and  dynamics 
of the system when the volume fraction increases. The number of particles corresponding to each volume fractions are tabulated below

\begin{center}
\begin{tabular} { | p {1.5cm}| p {1.5cm} | p {1.5cm}| p {1.5cm} |}
\hline
$\phi$ & $N$ & $n_s$ & $n_l$ \\
\hline
0.20 & 1055 & 938 & 117 \\
0.25 & 1320 & 1173 & 147 \\
0.30 & 1583 & 1407 & 176 \\
0.35 & 1847 & 1642 & 205 \\
0.40 & 2112 & 1877 & 235 \\
0.425 & 2243 & 1994 & 249 \\
0.45 & 2375 & 2111 & 264 \\
0.50 & 2639 & 2346 & 293 \\
\hline
\end{tabular}
\end{center}

\noindent where $N$, $n_s$, $n_l$ are the total number of particles, number of smaller particles and number of larger particles respectively. 
A series of simulations are performed for 8 different temperatures in the range $T \in (2.00, 0.25)$ for each total volume fraction. Each simulation 
runs for a total $5\times10^6$ steps for volume fraction upto $\phi=0.40$ with a time step of $dt=0.001$. For higher volume fractions
we have increased the length of simulation to $1\times10^7$ steps with a timestep of $dt=0.005$ for better statistics. It is repeated for three times for 
each state point and the average value of dynamical properties are calculated after 
the equilibration of the system.

\begin{figure}
\centering
\includegraphics[width=6cm, angle=-0]{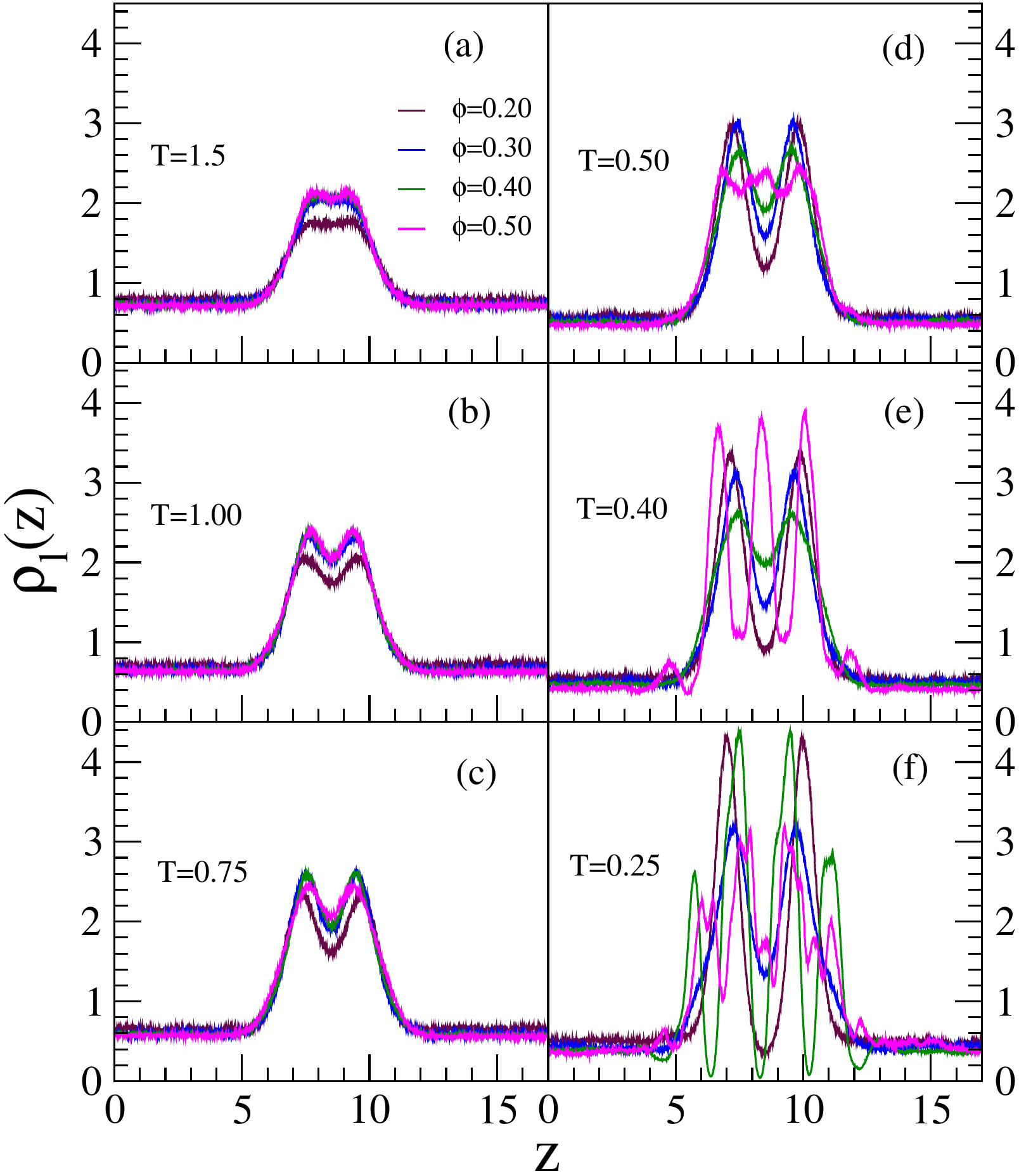}
\caption{Density profile of larger particles at four different volume fractions $\phi$ = 0.20, 0.30, 0.40 and 0.50 and at different temperatures (a) $T$=1.5, (b) $T$ = 1.0, (c) $T$ = 0.75, (d) $T$ = 0.50, (e) $T$ = 0.40 and (e) $T$ = 0.25.}
\label{denlarge}
\end{figure}

\section{Results and discussion}

\subsection{Spatial distributions and crystallization}

 It has been shown earlier that this model system at low volume fractions exhibits interesting structural properties due to the depletion interaction between the external potential barrier and larger particles\cite{anil}. The effective attraction between the potential barrier and larger particles results in an increased density of larger particles near the barrier, suggesting that a demixing can occur and the local concentration of larger particles near the barrier can be increased above the threshold for crystallization. Therefore,
 we systematically investigated the structural properties of the binary mixture at different total volume fractions as well as at different temperatures. 
 Please note that we simulated equi-volume binary mixtures at all volume fractions. Density profiles of smaller and larger 
 particles at different volume fractions and temperatures are shown in figures \ref{densmall} and \ref{denlarge} respectively. The density profile of smaller particles shows a minimum at the region of the external repulsive potential, as expected. This minimum in the density profile becomes sharper and sharper as we decrease the temperature. 
 As evident from figure \ref{densmall}, there are no significant changes in the density profile of smaller colloids as we increase the volume fraction of the mixture. However, the density profile of larger particles shows interesting changes with the decrease in temperature as well as with the increase in volume fraction.
 It is clear from the figure \ref{denlarge} that the larger particles get attracted to the region of the potential barrier at all temperatures, as indicated by the peak(s) in the region of the potential barrier. This is in agreement with earlier results for lower volume fractions\cite{anil1}. 
 At high temperatures, the normalized density profile does not show any significant changes with respect to the changes in volume fractions, except 
 that the height of the peak increases with volume fraction.
 However, as we decrease the temperature, the density profile starts developing multiple peaks in the region of potential barrier at higher volume fractions. 
 This indicates the layering of larger particles 
 in the region of the external potential barrier. This layering increases as we increase the volume fraction at low temperatures. The density profiles of smaller and larger particles clearly show that demixing occurs because of the depletion interaction between the potential barrier and larger particles. It also indicates that this phase separation becomes stronger as we decrease the temperature and increase the volume fraction. 
 
\begin{figure}
\centering
\includegraphics[width=6cm, angle=-0]{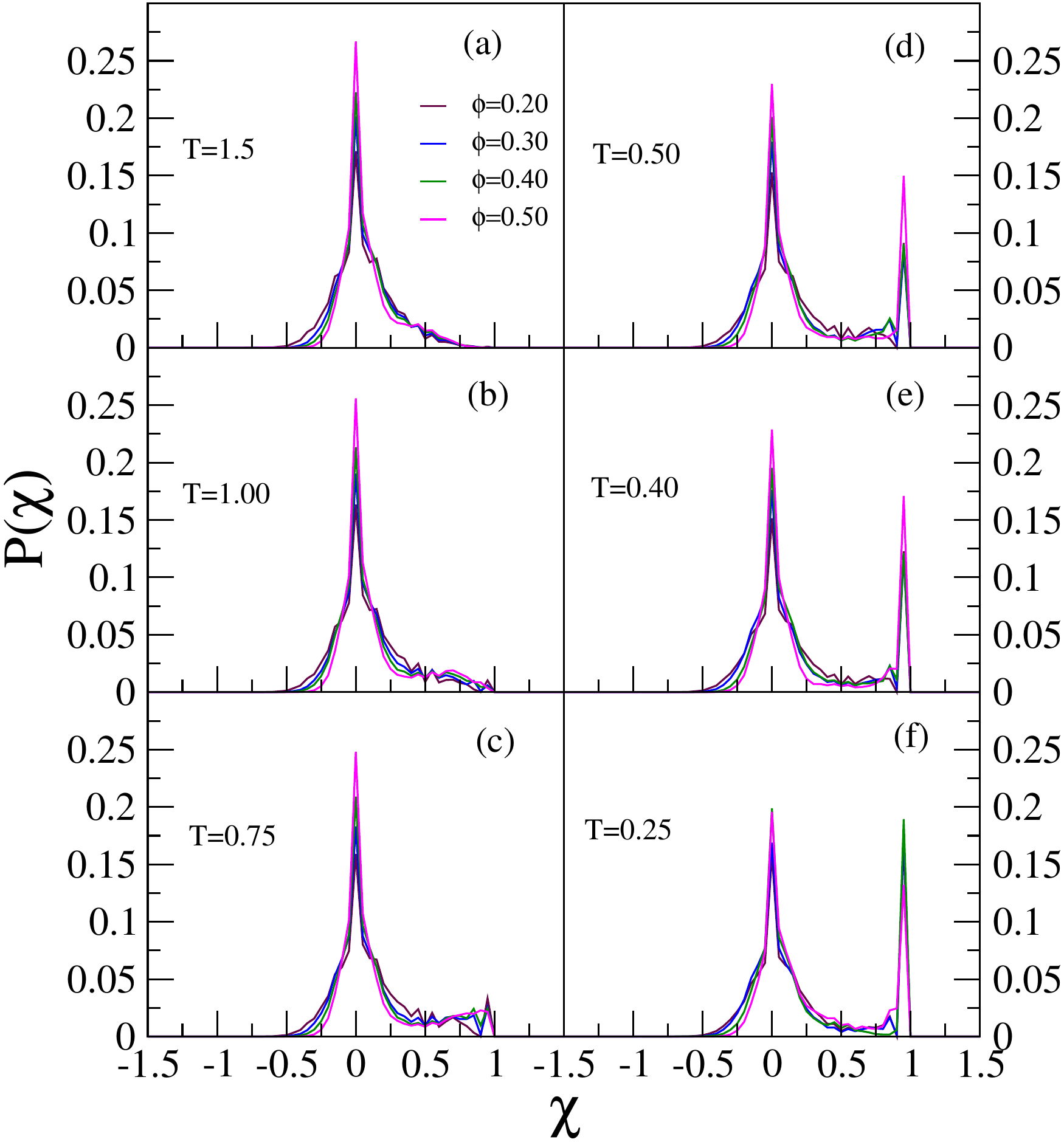}
\caption{Distribution, $P(\chi)$ of difference between the number of larger and smaller particles at different volume fractions and at different temperatures. Development of a peak at $\chi$ = 1 indicates the phase separation and formation of a region near potential barrier predominantly of larger particles. }
\label{pchi}
\end{figure}
 
 To quantify the extent of phase separation, we divided the simulation box into a number of rectangular boxes along the $z$-direction and calculated the difference in 
 the number of large 
 and small particles in each of the rectangular boxes. This difference in the number of large and small particles in the $i^{th}$ box is given by\cite{jalim,chari}
 
 \begin{equation}
  \chi = \frac{n_l^i - n_s^i}{n_l^i + n_s^i}
 \end{equation}
 
\begin{figure}
\centering
\includegraphics[width=8 cm, angle=-0]{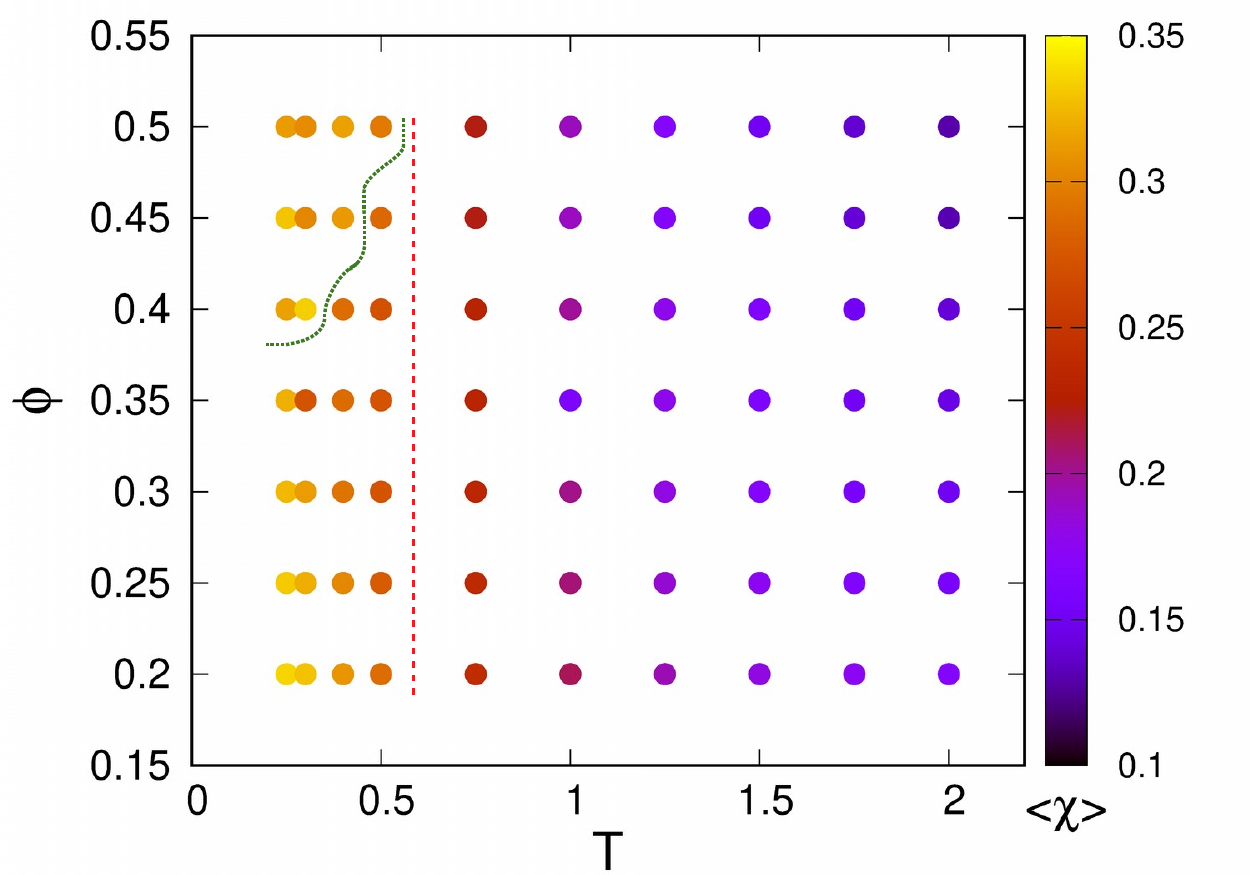}
\vspace{0.5cm}
\caption{$\phi \, \, vs. T$ phase diagram of the model system. The dashed line separates out the mixed and phase separated configurations and the dotted line separates out the fluid and crystal configurations.}
\label{phased}
\end{figure}
 
 Figure \ref{pchi} shows the distribution $P(\chi)$ of  $\chi$ at different volume fractions and at different temperatures. At high temperatures, $P(\chi)$ shows a single peak around 
 $\chi$ = 0 at all volume fractions, which shows the demixing is negligible. As the temperature decreases to 0.50, the height of this peak starts decreasing. The decrease in the peak 
 height of $P(\chi)$ at $\chi$ = 0 is accompanied by the appearance of another peak at $\chi$ = 1.0 at higher volume fractions. The appearance of this peak at $\chi$ = 1.0 means that larger particles are getting 
 accumulated in the region of the external potential barrier. The height of this second peak increases with decreasing temperature as well as increasing density or volume fraction.
 This indicates that the system phase separates into two: a mixed phase with both large and small particles and another phase with mostly large particles. This single component phase with large particles forms 
 in the region of the external potential barrier. 
  We have defined the average of $\chi$ as\cite{chari}
\begin{equation}
 \Phi(T,\phi) = \frac{1}{N_{box}}\Big<\frac{|n_l^i - n_s^i|}{n_l^i + n_s^i}\Big>
\end{equation}
where $N_{box}$ is the total number of rectangular boxes, the simulation box is divided into. We consider the value
of $\Phi(T,\phi)$ at which the peak at $\chi$ = 1 starts developing as the onset value of phase separation. Figure \ref{phased} shows the values of $\Phi(T,\phi)$ at different temperatures and volume fractions. At higher volume fractions and low temperatures, we get two phases where one of the phases is dominated by the larger particles and the other is a mixed phase with both large and small particles. The dashed line in the figure \ref{phased} separates these two phases. 

\begin{figure}
\centering
\includegraphics[width=7.5cm, angle=-0]{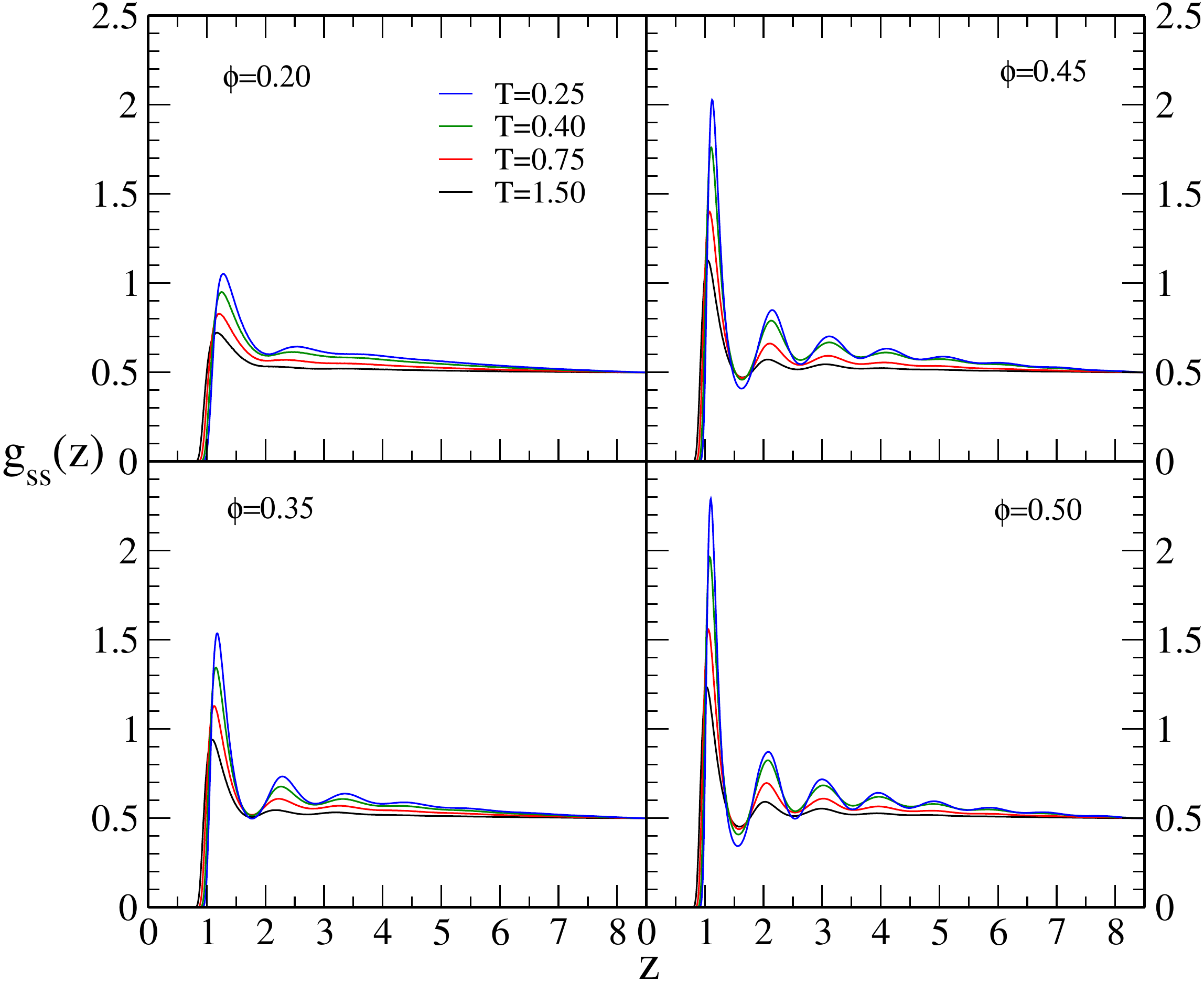}
\caption{ Radial distribution function of smaller particles for $\phi=0.20, 0.35, 0.45, 0.50$ and for T=0.25, 0.40, 0.75, 1.50.}
\label{rdfs}
\end{figure}

\begin{figure}
\centering
\includegraphics[width=7.5cm, angle=-0]{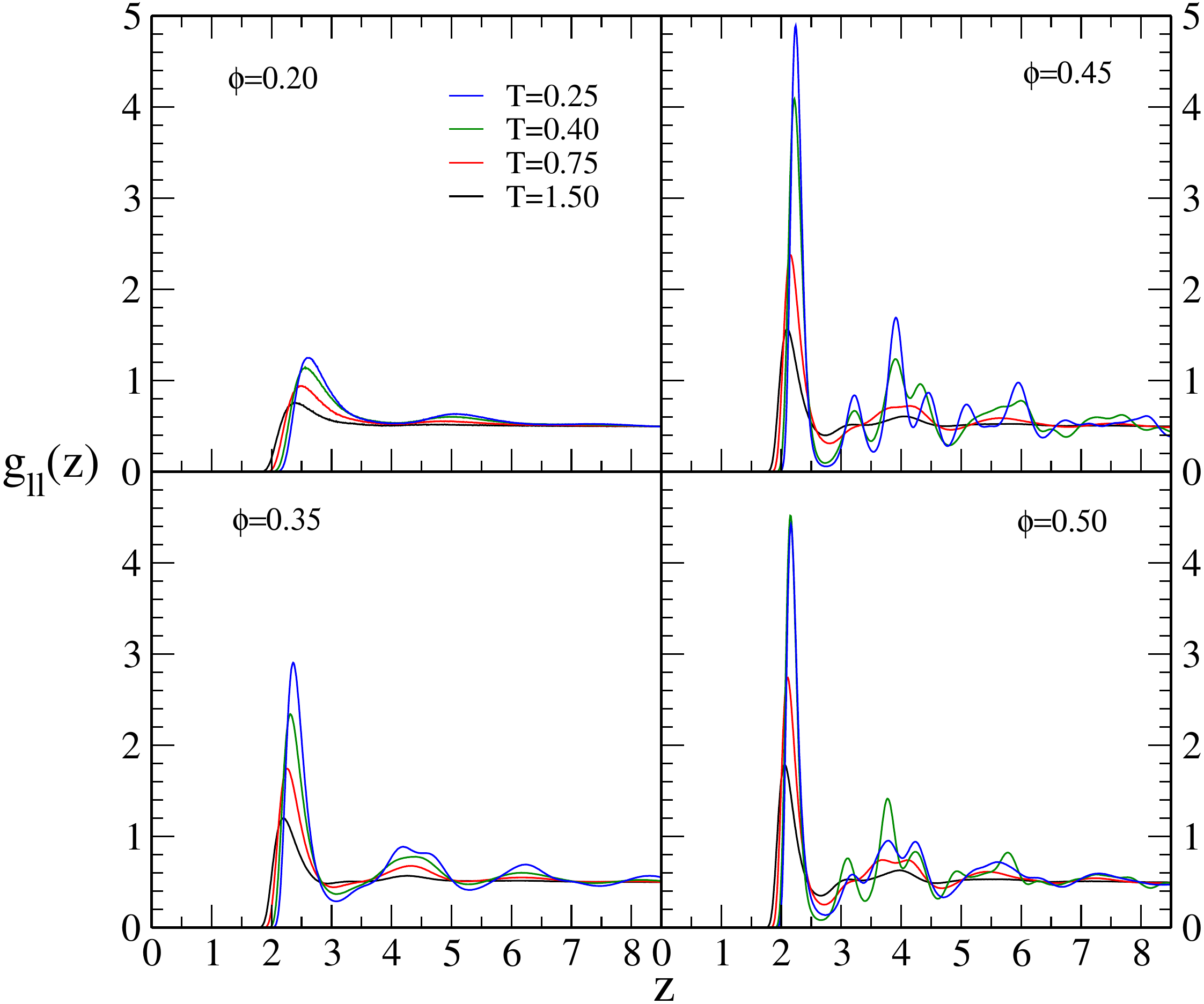}
\caption{ Radial distribution function of larger particles for $\phi=0.20, 0.35, 0.45, 0.50$ and for T=0.25, 0.40, 0.75, 1.50. The peaks in rdf at high volume fractions and low temperatures corresponds to an fcc lattice.}
\label{rdfl}
\end{figure}

 Since the density profile and $P(\chi)$ show that there is demixing at higher volume fractions and at low temperatures, it will be interesting to look at the structure of larger particles in the layers formed in the region of the external potential barrier. We have calculated the radial distribution function $g(r)$ of larger particles at different volume fractions and different temperatures, which is shown in figure \ref{rdfl}. For comparison, we have also plotted the radial distribution function of smaller particles, which is depicted in figure \ref{rdfs}. The $g(r)$ of smaller particles shows that they are in the liquid state at all volume fractions and temperatures. As the volume fraction increases, more layering occurs around the particles, which is expected. However, the $g(r)$ of larger particles changes significantly at higher volume fractions. At low volume fractions, larger particles are in a liquid state at all temperatures. As we increase the volume fraction to 0.35, they undergo crowding at low temperatures as indicated by the split in the second peak in $g(r)$. This is because more and more large particles are getting accumulated in the region of the external potential barrier. Further increase in the volume fraction results in the development of several peaks in $g(r)$ at low temperatures. This indicates that the crystallization of larger particles in the central region is due to the depletion interaction. This is interesting since the volume fraction of large particles is much lower than the volume fraction (~ 20 \%) required for crystallization in single component systems. 
 This occurs since the local volume fraction of larger particles is high enough to form crystals. The position of the peaks in the $g(r)$
 indicates that the crystal formed is of face centered cubic structure. Our results suggest that crystallization can be achieved in low volume fractions by making use of depletion interaction between the larger particles and external repulsive potential. 
 Earlier, crystallization has been achieved for colloidal systems with long range repulsions at low volume fractions(Wigner crystals). 
 In our model system, the interactions between the particles are of the form $r^{-12}$, which is a much shorter range. This essentially means that the volume 
 fraction required for crystallization can be brought down by making use of the depletion interaction between the external potentials and larger particles to manipulate the local volume fraction. The volume fraction of the larger particles to crystallize  may be brought down further by increasing the strength of depletion interaction.

 \subsection{Dynamics}
 
\begin{figure}
\centering
\includegraphics[width=8 cm, angle=-0]{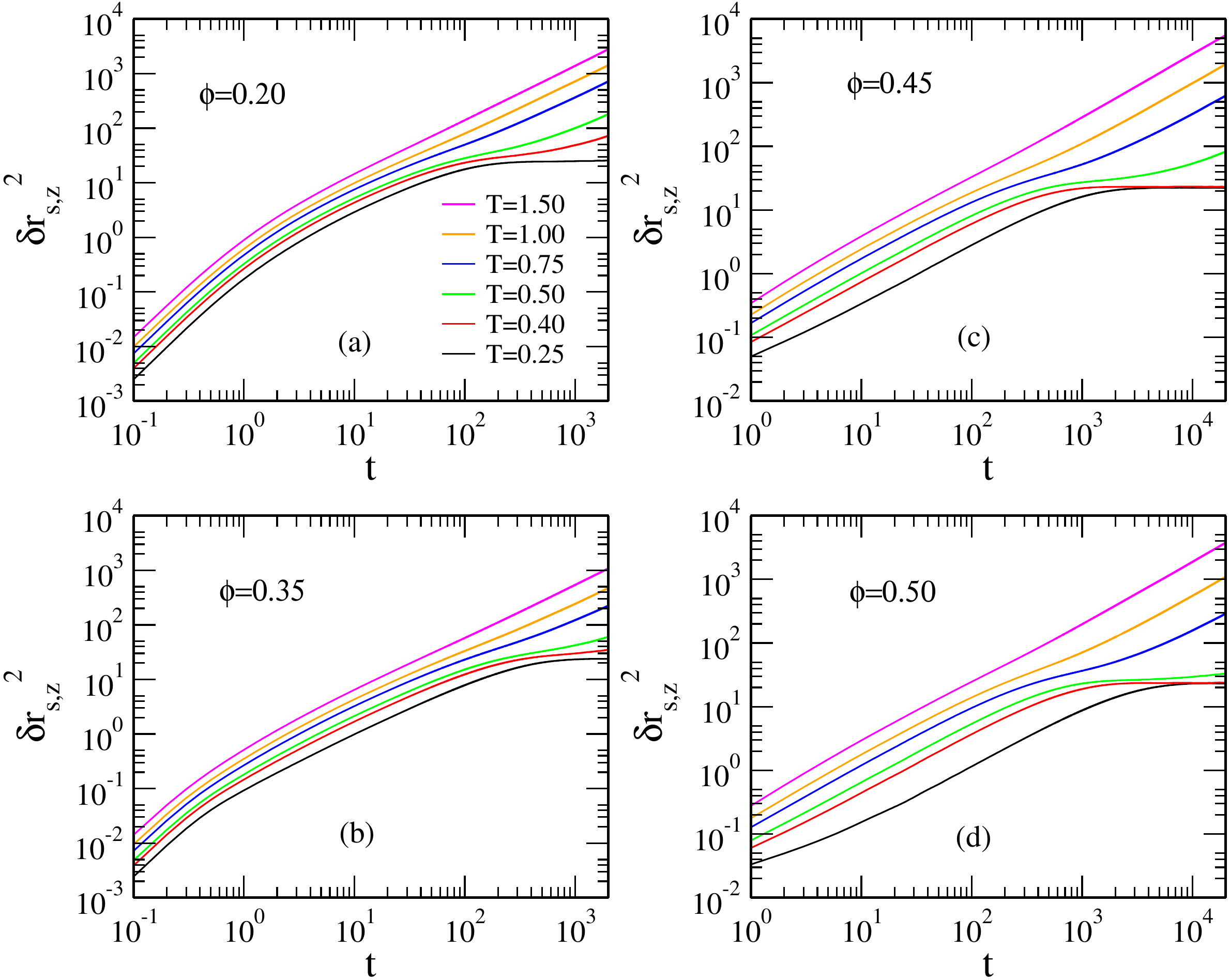}
\caption{Mean squared displacement of smaller particles along the $z$-direction for $\phi=0.20, 0.35, 0.45, 0.50$ and for T=0.25, 0.4, 0.5, 0.75, 1.0 and  1.5. At low 
temperatures, MSD shows a plateau at larger times.}
\label{msdsz}
\end{figure}

  In order to look at the changes in the dynamics of smaller and larger particles when demixing and subsequent crystallization happen, we have calculated different dynamical properties of the binary colloidal system.
  It has been shown earlier that this model at low volume fractions shows intriguing dynamical properties of both the smaller and the larger particles. For example, the dynamics of smaller particles at a lower temperature are similar to that of supercooled liquids, even though the volume fractions are very small\cite{anil1}.
  This is manifested by the appearance of a plateau in the mean squared displacement at intermediate times, stretched exponential decay of intermediate scattering function, nonzero values of non-Gaussian parameters, etc.  Figure \ref{msdsz} shows the mean squared displacement(MSD) 
  of smaller particles along $z$-direction for different temperatures at four different volume fractions. At higher temperatures, the MSD is ballistic at earlier times, crossing over to a diffusive regime at longer times. As we decrease the temperature, the dynamics of smaller particles slow down, 
  developing a plateau at intermediate times before crossing over to diffusive dynamics. These results are in agreement with earlier investigations at 
  low volume fractions\cite{anil1,mustakim}. However, at high volume fractions and low temperatures, MSD reaches the plateau and continues to stay there even at very large times. This is related to the crystal formation of larger particles in the region of the external potential barrier. Once the larger particles form a compact crystal, the smaller particles find it impossible to penetrate through this structure and remain completely localized between two crystalline domains(crystalline domains are repeated along the $z$-direction because of the periodic boundary conditions). This complete localization of smaller particles is reflected in the saturated value of mean squared displacement.
  
\begin{figure}
\centering
\includegraphics[width=8cm, angle=-0]{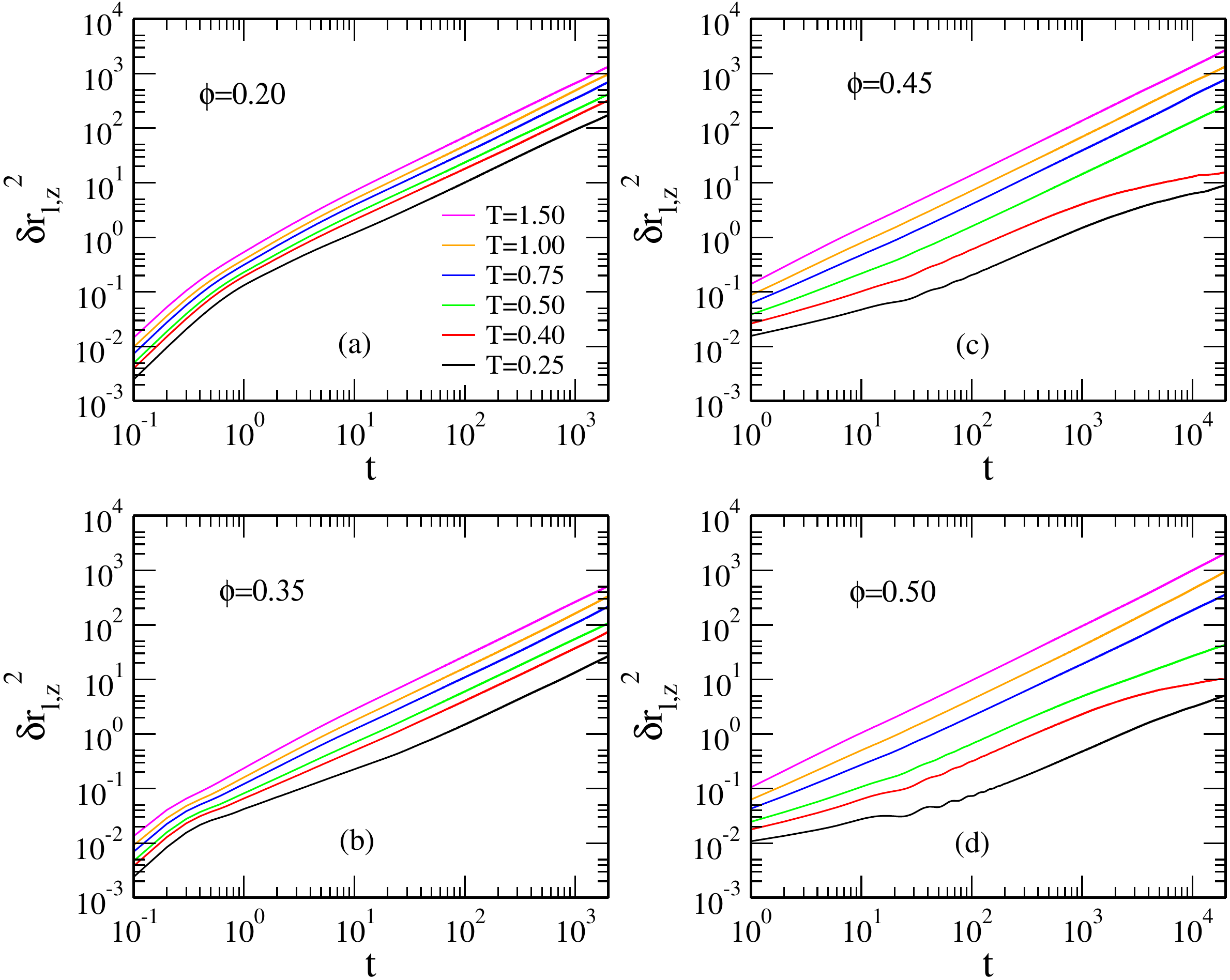}
\caption{Mean squared displacement of larger particles along the $z$-direction for $\phi=0.20, 0.35, 0.45, 0.50$ and for T=0.25, 0.4, 0.5, 0.75, 1.0 and  1.5. MSD is linear
except for those phase points where crystalline formation occurs.}
\label{msdlz}
\end{figure}

\begin{figure}
\centering
\includegraphics[width=7.5 cm, angle=-0]{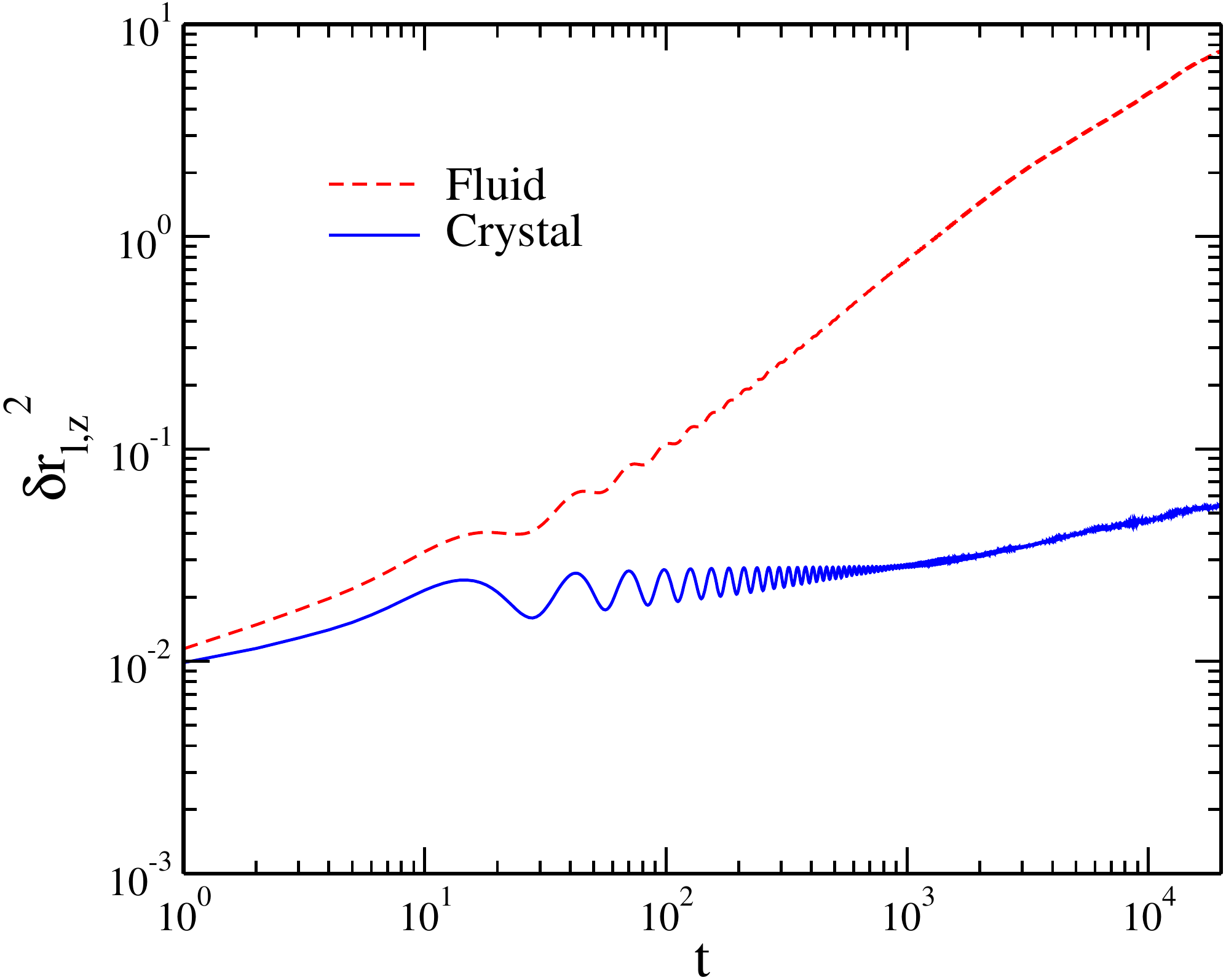}
\caption{Mean squared displacement of larger particles which are in the crystalline phase(solid line) and fluid phase(dashed line) at $\phi$=0.50 and $T$=0.25 along $z-$direction. MSD's at other phase points where a crystalline domain present are similar.}
\label{msdlzcnc}
\end{figure}

The dynamics of larger particles at low volume fractions is found to be completely contrasting compared to that of smaller particles. Because of 
the depletion interaction between the external potential barrier and larger particles, MSD of larger particles remains linear at very low temperatures, 
thus does not show any slowing down of dynamics. We found that our results are matching with these results till the total volume fraction is 0.4. 
Figure \ref{msdlz} depicts the MSD of larger particles along $z$-direction at different volume fractions and at different temperatures. As evident from figure \ref{msdlz}, at volume fractions larger than 0.4, the mean squared displacement of larger particles deviates from the linear behavior and shows signs of slowing down at lower temperatures. Here again, the demixing and formation of the crystalline phase cause this slowing down behavior of dynamics. 
It will be interesting to observe the dynamics of larger particles that are in the crystal phase and in the fluid phase separately. So we have separated out the mean squared displacement  of larger particles that are participating in the crystal formation and that are in the fluid phase along the $z$-direction. 
This is shown in figure \ref{msdlzcnc} for one phase point ($\phi$ = 0.50 and $T$ = 0.25). We get similar graphs for other phase points where a crystalline domain is present. The mean squared displacement of particles in the crystalline environment reaches a plateau within a very short time, which is as expected.
Please note that a very small increase in MSD occurs at very large times. This is due to the surface atoms which get detached from the crystalline phase and enter into the fluid region. However, we found that such occurrences are very small in number and do not alter the average value of MSD significantly.
The MSD of larger particles in the fluid phase increases with time linearly. The fluctuations from the linear behavior in their MSD is due to the fact that the number of larger particles in the fluid phase is quite small leading to poor statistics.

\begin{figure}
\centering
\includegraphics[width=8 cm, angle=-0]{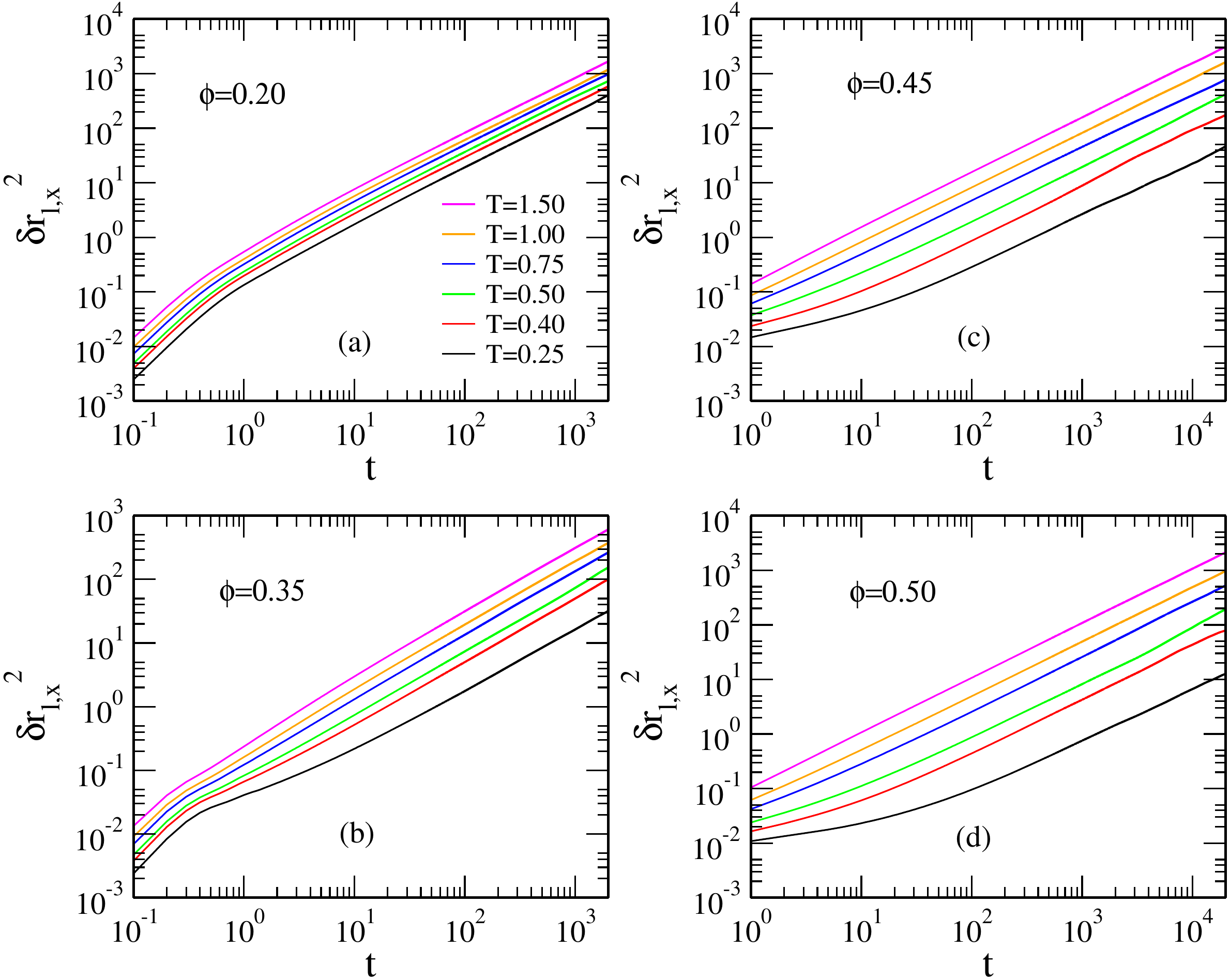}
\caption{Mean squared displacement of larger particles perpendicular to the external potential barrier for $\phi=0.20, 0.35, 0.45, 0.50$ and for T=0.25, 0.4, 0.5, 0.75, 1.0 and  1.5.}
\label{msdlx}
\end{figure}

\begin{figure}
\centering
\includegraphics[width=7.5cm, angle=-0]{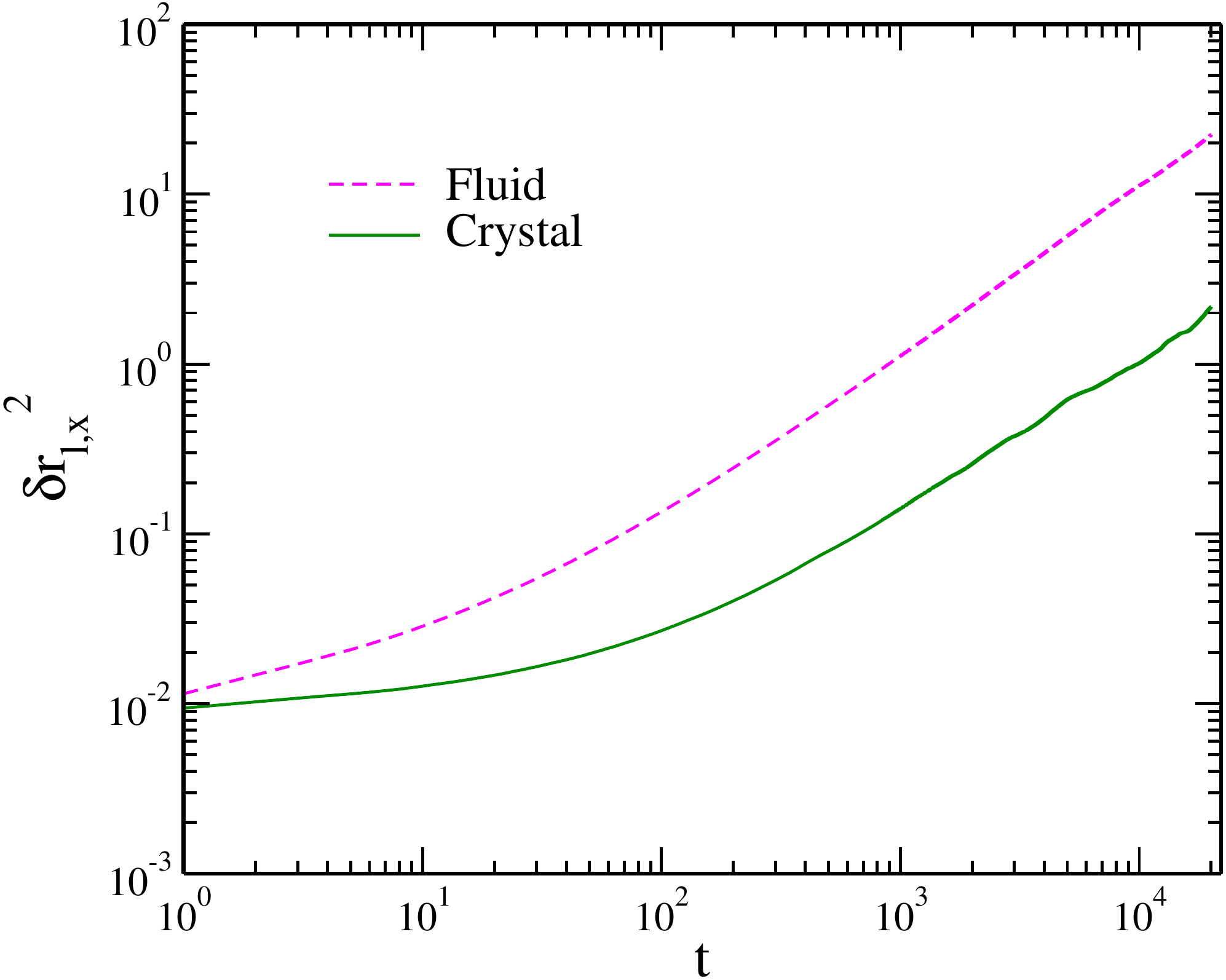}
\caption{Mean squared displacement of larger particles perpendicular to the external potential barrier which are in the crystalline phase(solid line) and fluid phase(dashed line) at $\phi$=0.50 and $T$=0.25. MSD's at other phase points where a crystalline domain present are similar.}
\label{msdlxcnc}
\end{figure}

We have also investigated the dynamics of larger particles perpendicular to the direction of external potential barrier. Since particles do not have to encounter the external 
potential barrier in these directions, there will not be any direct effect of depletion interaction in their dynamics in these directions, i.e., the consequence of depletion interactions on dynamics in these directions occurs via the demixing and subsequent crystallization of the system. In figure \ref{msdlx}, the MSD of larger particles along the perpendicular direction of the external potential barrier is given.
As expected, the MSD is linear in time and does not show any signatures of slowing down. 
As in the case above, we separated out the MSD's of larger particles in the crystalline and fluid environment separately. This is plotted  
for the phase point($\phi$ = 0.50 and $T$ = 0.25) in figure \ref{msdlxcnc}. The MSD of larger particles in the fluid phase increases with time linearly as expected. However, interesting behavior occurs in the dynamics of larger particles in the crystal phase. Their MSD also increases linearly with time and shows a nonzero diffusion coefficient. This shows that while the crystalline domain remains intact in its structure, it diffuses as a whole in the direction perpendicular to the external potential barrier (see figure \ref{msdlxcnc}). Such moving crystals have been reported before, but predominantly in nonequilibrium systems where the crystals are exposed to heat or light\cite{panda1,panda2,shen,commins}. 
Dynamic crystals have also been reported in the case of active colloids\cite{shen}. 
However, our model system is an equilibrium system
with an external potential involved. Recently, a dynamical crystal phase has been reported in a soft crystal with interstitial dopants near their melting point\cite{tauber}. Although, here the body centered cubic structure of the soft crystal remains intact on average, but the instantaneous structure shows significant deviations from the perfect lattice. We have not observed such larger deviations in our system. This warrants further investigation on this diffusive behavior of the crystalline domain to ascertain the physical reasons behind it. This will be done in due time and, the results will be reported elsewhere.

\begin{figure}
\centering
\includegraphics[width=7.5cm, angle=-0]{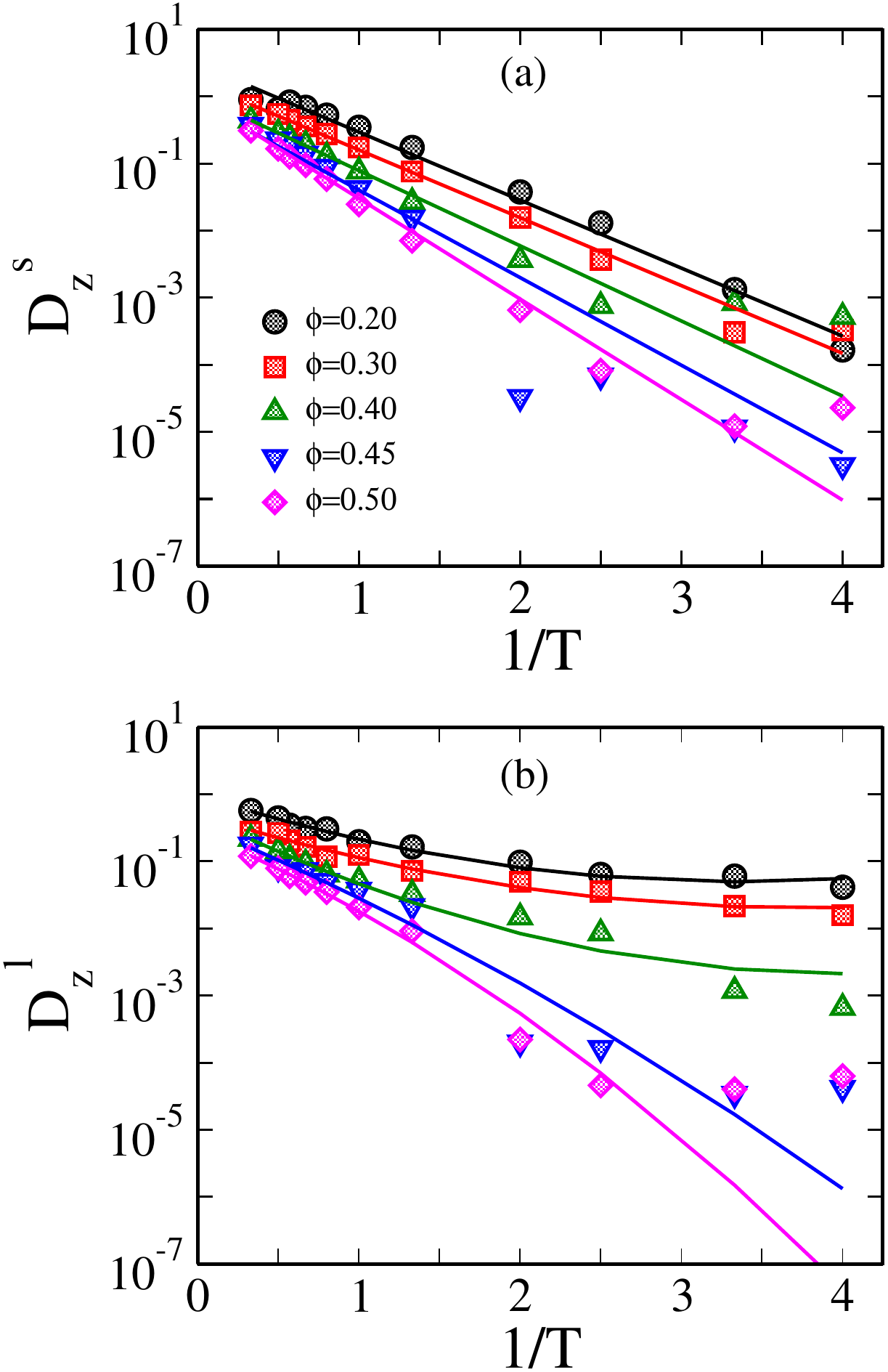}
\caption{$log(D_z)$ versus inverse temperature curve for (a)smaller and (b)larger particles at different volume fractions. The points are from simulations and the curve is a fit to the (a) Arrhenius and (b) $d$-Arrhenius equation(eqn. 5)}
\label{arrfit}
\end{figure}

\begin{figure}
\centering
\includegraphics[width=7cm, angle=-0]{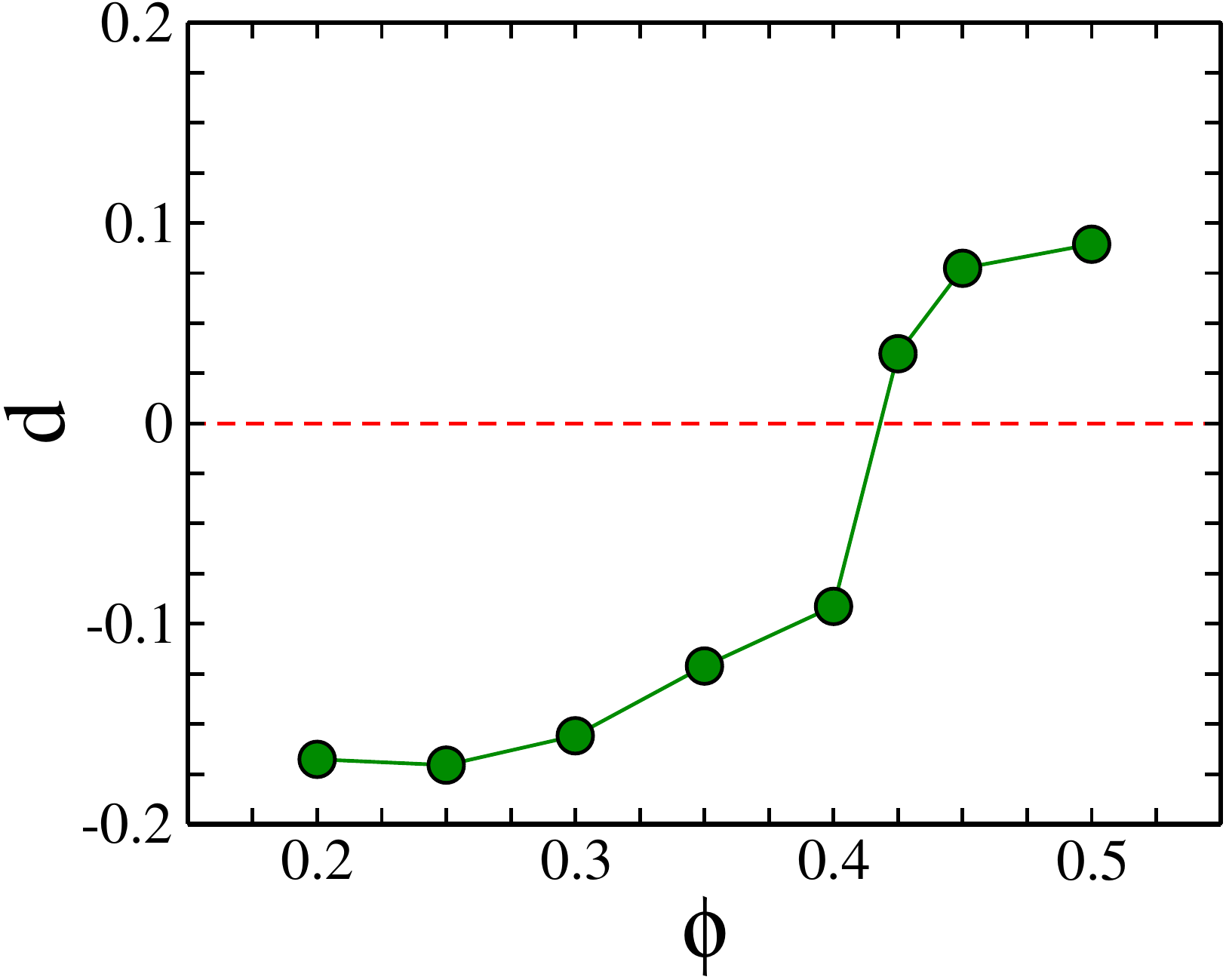}
\caption{The deformation parameter $d$ versus the  total volume fraction $\phi$. The sign of $d$ changes from negative to positive at $\phi$ = 0.425, 
indicating the crossover from sub-Arrhenius to super-Arrhenius diffusion.}
\label{dvalue}
\end{figure}

The temperature dependence of the self diffusion coefficient of fluids usually follows an Arrhenius behavior\cite{Arrhenius,Kramers}. However, deviations from Arrhenius behavior are reported in many cases, especially at lower temperatures. Aquilanti and coworkers proposed a formalism based on Tsallis' nonextensive statistical mechanics\cite{tsallis} to include these deviations also into the framework of Arrhenius equation\cite{aquilanti1,aquilanti2,silva1,silva2,nishiyama1}. They have proposed a deformed Arrhenius equation
as
\begin{equation}
 D(T) = A\Big[1-d\frac{E_0}{k_BT}\Big]^{1/d}
\end{equation}

\noindent which includes Arrhenius behavior as well as the deviations from it. Here $E_0$ is the height of the potential barrier. $d$ is the parameter, known as 'deformation parameter', which will determine the behavior of $D$ with respect to temperature. If the diffusion is Arrhenius in nature, then $d=0$. For positive values of $d$, the temperature 
dependence of diffusion is termed as 'super-Arrhenius', where we get convex curve for $log(D) \, \, vs. 1/T$. When the $log(D) \, \, vs. 1/T$ curve is concave, the value of $d$ is negative and diffusion is said to be 'sub-Arrhenius'.
In most of the observed deviations, super-Arrhenius behavior occurs in classical systems where correlated 
motion plays an important role as in the case of supercooled liquids\cite{Stirnemann, desouza}. Sub-Arrhenius behavior is mostly found in processes where quantum effects 
play an important role as in the case of tunneling effect during chemical reactions\cite{aquilanti1, liang, braun}. However,
it has been shown that self-diffusion of larger particles in a binary mixture subjected to an external potential barrier at low volume fractions is found to be 
sub-Arrhenius in nature with a negative $d$ parameter\cite{mustakim}. Hence, we have investigated the temperature dependence of self-diffusion of 
larger particles at different volume fractions. Figure \ref{arrfit} shows the $log(D_z)$ versus inverse temperature curve for different volume fractions. 
It is evident from the figure that at low volume fractions  $log(D_z) \, \, vs. 1/T$ curve is concave and the diffusion of large particles follows a sub-Arrhenius behavior.
However, as we increase the volume fraction, the concavity of the $log(D_z) \, \, vs. 1/T$ curve decreases and, above $\phi$ = 0.4, the curve becomes convex. This means that 
at higher volume fractions, there is a cross over from sub-Arrhenius
to super-Arrhenius behavior in the diffusion of larger particles. To make it more quantitative, we have fitted the $log(D_z) \, \, vs. 1/T$ curves with the deformed Arrhenius equation and extracted the values of $d$ parameter for different volume fractions. These are plotted in figure \ref{dvalue}. At low volume fractions, $d$ is negative indicating the sub-Arrhenius nature. At $\phi$ = 0.425, $d$ suddenly changes the sign and becomes positive, indicating a cross over to super-Arrhenius behavior. 
It should be noted that this crossover coincides with the crystallization of larger particles in the region of the potential barrier. As we have seen earlier, when 
the crystallization happens in the barrier region, most of the larger particles participate in crystal formation and, the number of larger particles in the fluid region becomes much smaller. So the overall diffusion of larger particles becomes suppressed and this is manifested as the super-Arrhenius behavior.

\section{Conclusions}

 We have investigated the demixing of a binary colloidal mixture subjected to an external potential barrier. The attractive depletion interaction between the potential barrier and larger particles in the binary mixture causes it to phase separate into two regions: one a pure phase comprised of only larger particles in the 
 region of the potential barrier and another a mixed phase elsewhere. This phase separation allows one to manipulate the local volume fraction of large particles high enough 
 that they undergo crystallization into an fcc structure. In general, such manipulations of local density are done by nonequilibrium processes like dielectrophoresis 
 and diffusiophoresis. However, the binary colloidal mixture we have investigated is in equilibrium and, the changes in the local density of colloidal particles are done through the depletion interaction. The crystal domain, formed in the region of the external potential barrier, moves perpendicular to the external potential. Again such moving crystals are generally observed in nonequilibrium systems, where the 
 crystallites are subjected to driving forces. Our model system does not have any such driving forces and the physical reasons behind these dynamics is unclear. 
 It will be interesting to investigate this moving crystalline domain further. Work in this direction will be taken up and, results will be reported in due time. 
 The formation of this crystalline phase coincides with a change over from sub-Arrhenius to super-Arrhenius diffusion of larger particles. It has been concurred that sub-Arrhenius behavior occurs in processes where quantum effects such as tunneling are predominant while super-Arrhenius diffusion occurs in classical systems where
 particles undergo correlated dynamics such as in supercooled liquids. However, our model system of binary colloids is purely a classical system, where both types of behaviors occur. At low densities, the diffusion is sub-Arrhenius because the depletion attraction between the barrier and the larger particles effectively reduces
 the activation energy for diffusion and makes it temperature dependent. At higher densities, depletion interaction causes phase separation and subsequent crystallization of larger
 particles, hindering their diffusion. Thus the diffusive behavior changes from sub-Arrhenius to super-Arrhenius.

The authors acknowledge the financial support from Department of Atomic Energy, India through the 12th plan project(12-R\&D-NIS-5.02-0100)

\bibliography{aps} 
\bibliographystyle{aps} 

\end{document}